\documentclass{emulateapj}

\slugcomment{In press on ApJ}

\newcommand{\ltsima} {$\; \buildrel < \over \sim \;$}
\newcommand{\gtsima} {$\; \buildrel > \over \sim \;$}
\newcommand{\lta} {\lower.5ex\hbox{\ltsima}}
\newcommand{\gta} {\lower.5ex\hbox{\gtsima}}

\shorttitle{Infrared jet in 3C~401}
\shortauthors{Chiaberge et al.}

\newcommand\gtsim{\mathrel{\lower0.6ex\hbox{$\buildrel {\textstyle >}
      \over {\scriptstyle \sim}$}}}
\newcommand\ltsim{\mathrel{\lower0.6ex\hbox{$\buildrel {\textstyle <}
      \over {\scriptstyle \sim}$}}}

\begin{document}

\title{The infrared-dominated jet of 3C~401}

\author{Marco Chiaberge\altaffilmark{1},William.~B. Sparks, F.~Duccio Macchetto\altaffilmark{2}}\affil{Space Telescope Science Institute, 3700 San Martin Dr., Baltimore, MD 21218}
\email{chiab@stsci.edu}
\author{Eric Perlman}
\affil{Joint Center for Astrophysics, Physics Department, University of Maryland, Baltimore County, 1000 Hilltop Circle, Baltimore, MD 21250}
\author{Alessandro Capetti}
\affil{INAF - Osservatorio Astronomico di Torino, Strada Osservatorio, 20 - I 10025 Pino Torinese}
\author{Barbara Balmaverde}
\affil{Universit\`a di Torino, Via P. Giuria 1, I 10125 Torino}
\author{David Floyd}
\affil{STScI, 3700 San Martin Dr., Baltimore, MD 21218}
\author{Christopher O'Dea, David J. Axon}
\affil{Department of Physics, Rochester Institute of Technology, 85 Lomb Memorial Drive, Rochester, NY 14623}

\altaffiltext{1}{On leave from INAF-Istituto di Radioastronomia, Via P. Gobetti 101, Bologna, Italy, 40129-I}
\altaffiltext{2}{On assignment from ESA}

\begin{abstract}

We present  a Hubble Space Telescope  image of the  FR~II radio galaxy
3C~401,  obtained at  1.6$\mu$m with  the  NICMOS camera  in which  we
identify the infrared counterpart of the brightest region of the radio
jet. The jet has a complex radio structure and brightens where bending
occurs, most likely  as a result of relativistic  beaming.  We analyze
archival data in the radio, optical  and X-ray bands and we derive its
spectral energy distribution.  Differently  from all of the previously
known optical extragalactic jets, the jet in 3C~401 is not detected in
the  X-rays even  in  a long  48ksec  X-ray Chandra  exposure and  the
infrared  emission dominates  the overall  SED.  We  propose  that the
dominant  radiation mechanism  of this  jet is  synchrotron.   The low
X-ray emission is then caused by two different effects: i) the lack of
any strong  external photon  field and ii)  the shape of  the electron
distribution.  This  affects the location  of the synchrotron  peak in
the  SED, resulting  in a  sharp cut-off  at energies  lower  than the
X-rays.  Thus  3C~401 shows a new  type of jet  which has intermediate
spectral  properties between  those of  FR~I, which  are  dominated by
synchrotron emission up to X-ray  energies and FR~II/QSO, which show a
strong  high-energy  emission  due  to inverse-Compton  scattering  of
external  photons.   This  might be  a  clue  for  the presence  of  a
continuous  ``sequence'' in  the properties  of large  scale  jets, in
analogy with the ``blazar sequence'' already proposed for sub-pc scale
jets.

\end{abstract}



\section{Introduction}

The  study of  relativistic  jets, which  constitute a  distinguishing
feature  of  radio-loud AGN,  has  recently  been  revitalized by  the
discovery that a  large number of radio jets are  bright in the X-rays
\citep[e.g.][]{chartas,sambruna04,worrall01}.    While   the  emission
process  in  the  radio  band  is  certainly  non-thermal  synchrotron
radiation, the origin  for the high energy emission  is still unclear.
Synchrotron  and  inverse-Compton  radiation  are  the  most  accepted
interpretations.  The former process seems to predominate in low-power
jets  associated  with radio  galaxies  belonging  to  the FR~I  class
\cite{fr}, while  in powerful quasars and radio  galaxies the dominant
mechanism is most likely  inverse-Compton.  In powerful QSOs, the seed
photons  for  scattering  may  be  provided by  the  cosmic  microwave
background radiation \citep{tavecchio00,cgc01}.  Such a scenario needs
the jet  to be relativistic on large  scales (up to $\sim  100$ kpc or
more),  so that  the  CMB  photon field  is  enhanced by  relativistic
beaming  effects. This  requirement  is thus  considered  as a  strong
further  evidence for   relativistic motion  on  large scales,  in
addition  to  the well  known  jet  to  counter-jet asymmetries,  lobe
depolarization    asymmetries    \citep{laing88,garrington88},    core
enhancement  \citep[e.g.][]{giovannini94} and superluminal  motions in
QSO and  radio galaxies in which the  jet axis forms a  small angle to
the line-of-sight \citep{rees66,whitney71}.

Here we focus on 3C~401, which is a powerful FR~II radiogalaxy located
at a  redshift of $z=0.201$.  In  the radio, it shows  a one-sided jet
and  two hot-spots  which are  clearly  visible although  they do  not
appear  as  compact as  in  more  powerful  (and more  distant)  FR~II
\citep{hardcastle98}.  Nevertheless, its high radio-power ($L_{178MHz}
= 2 \times 10^{34}$ erg  s$^{-1}$ Hz$^{-1}$) is one order of magnitude
higher than  the fiducial  FR~I/FR~II break.  On  the other  hand, the
characteristics  of the  environment are  more typical  of  FR~I radio
galaxies, since  the object is the cD  of an Abell class  1 cluster of
galaxies, while FR~II are generally found in small groups.  In the HST
optical images \citep{dekoff} the host galaxy appears as an elliptical
galaxy of low surface brightness, with no clear sign of dust features.
A  close companion  galaxy  is located  at  $\sim 4$  arcsec from  the
nucleus of  the radio source.  From  the point of view  of its optical
spectrum, 3C~401 belongs to the  subclass of FR~II which are dominated
by  low excitation emission  lines (Low  Excitation Galaxy,  LEG). The
object  is again  more similar  to  FR~I also  as far  as the  nuclear
properties  are  concerned: the  host  galaxy  shows  a faint  optical
unresolved  core  which  is  most  likely interpreted  as  the  optical
counterpart  of the  synchrotron radio  core, similarly  to  the large
majority  of  3CR  FR~I   \citep{pap4}.   Therefore,  its  high  power
associated  with  FR~I-like nuclear  characteristics  makes 3C~401  an
extremely  interesting  object.   A  detailed  study  of  the  jet  in
``transition'' sources may shed  light on the physical processes which
determine the properties of jets in sources of different total power.

In  section  \ref{observations} we  describe  our HST/NICMOS  infrared
observations, in section \ref{morph} we outline the jet morphology and
we  perform photometry,  while observations  at other  wavelengths are
analyzed in section \ref{otherobs}. In section \ref{sedsect} we derive
and model the spectral energy  distribution of the jet, and in section
\ref{conclu} we discuss our findings and draw conclusions.

H$_0 =  75$ km s$^{-1}$ Mpc$^{-1}$  and q$_0=0.5$ are  used throughout
the paper.

\section{Infrared HST observations}
\label{observations}

We  have observed  3C~401  with  HST/NICMOS as  part  of our  snapshot
program GO~10173. Our complete sample  is derived from the 3CR catalog
of extragalactic  radio sources, limited to $z<0.3$  to take advantage
of the highest spatial resolution HST can offer.  The sample comprises
115 sources, 18 of which have been observed during previous cycles, as
part of other HST programs.

The observations of  3C~401 were performed on August  11 2004. We used
NICMOS  camera  2   and  the  F160W  filter,  which   is  centered  at
1.6037$\mu$m and covers the spectral  range from 1.4 to 1.8$\mu$m.  At
the redshift of the source, the Pa$\beta$ emission line is included in
the filter.  However, due to  its large bandwidth,  continuum emission
dominates.    The   field   of   view   of   the   NIC-2   camera   is
19.2$^{\prime\prime}  \times 19.2  ^{\prime\prime}$ and  the projected
pixel  size is  0.076$^{\prime\prime}  \times 0.075  ^{\prime\prime}$.
The total exposure time is 1152s, split into 4 images of 288s allowing
4-point dithering with sub-pixel spacing, in order to improve both the
PSF sampling and bad-pixel removal.  The details of the data reduction
will be given in a  forthcoming paper (Madrid et al., in preparation).
Here we only briefly describe the main steps of the reduction.

\begin{figure}
\epsscale{1.1}
\plotone{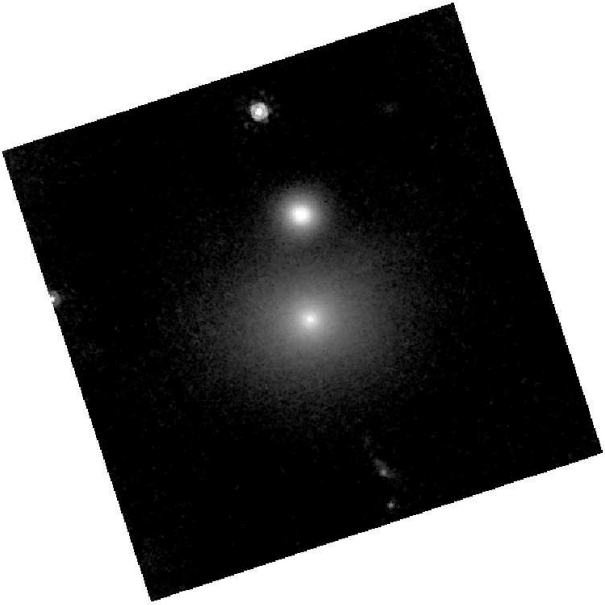}
\caption{NICMOS Image of 3c~401. North is top, East is left. The field
of view is $18^{\prime\prime}\times18^{\prime\prime}$.}
\label{final}
\end{figure}

The  raw data  have been  retrieved  from the  HST MAST  (Multimission
Archive at  Space Telescope) and  have been processed by  the standard
OTFR (On  The Fly Reprocessing)  calibration pipeline, which  uses the
latest calibration files.  The subsequent reduction has been performed
using   IRAF.    As  a   first   step,   we   remove  the   ``pedestal
effect''\footnote{The  pedestal effect  is one  of the  main anomalies
that appear in NICMOS images. It is a time varying, quadrant dependent
bias level which appears in  all NICMOS images.  For more details, see
http://www.stsci.edu/hst/nicmos.}  from each  of the images using {\it
pedsky},  and we  subtract  the residual  background  level.  We  then
combine the  four calibrated images using the  {\it drizzle} algorithm
\citep{drizzle}.    The   final  ``drizzled''   image   is  shown   in
Fig.~\ref{final}.  The projected  pixel  size in  the  final image  is
$0.038^{\prime\prime}$.

The  host galaxy of  the radio source 3C~401 is  the largest object at
the center of the field  of view.  A  faint point source is present at
the  center of the galaxy.   A close companion,  which appears to be a
smaller elliptical galaxy, is clearly visible, $\sim 4^{\prime\prime}$
north of the nucleus of our target. A bright unresolved source located
$\sim 8^{\prime\prime}$ north  of the galaxy is  present in our image.
This object   is  also visible in  optical   HST images, where  it  is
slightly resolved. Since it does not appear to be related to any radio
emitting feature, this object is most  likely an infrared-bright field
galaxy and will be studied elsewhere.

\begin{figure}
\epsscale{1}
\plotone{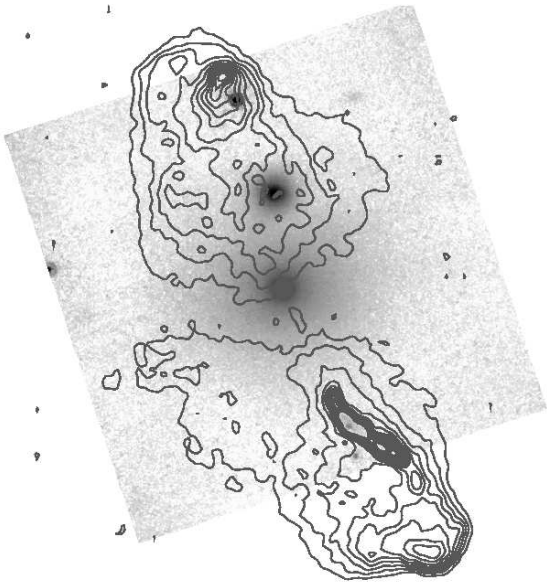}
\caption{Overlay of the radio contours onto our NICMOS image.}
\label{overlay}
\end{figure}

The most intriguing feature in the image is the  IR counterpart of the
radio jet,  located about 6  arcsec south-west  of  the nucleus, at  a
position angle  of $\sim 205^{\circ}$.  In  Fig. \ref{overlay} we show
the  overlay of the  radio  contours at  1.5  GHz onto our  HST/NICMOS
image.   Radio  data are   from VLA/MERLIN observations \citep{treichel}, and  have  an  angular  resolution of
0.35 arcsec (circular  beam).  We have used the  position of the radio
core as reference to align the  world coordinate system of  the HST
image to the radio data frame.

The brightest region  of the radio   jet, which appears to be  located
where  the jet  bends,  and its  infrared   counterpart clearly overlap.
However, it appears that the IR jet is detected over a significantly
smaller area than the radio jet.  In the following section we describe
the   IR jet  morphology   and  we outline  our    method for aperture
photometry of the components.

\section{The infrared jet in 3C~401: morphology and photometry}
\label{morph}

\begin{figure*}
\epsscale{1}
\plotone{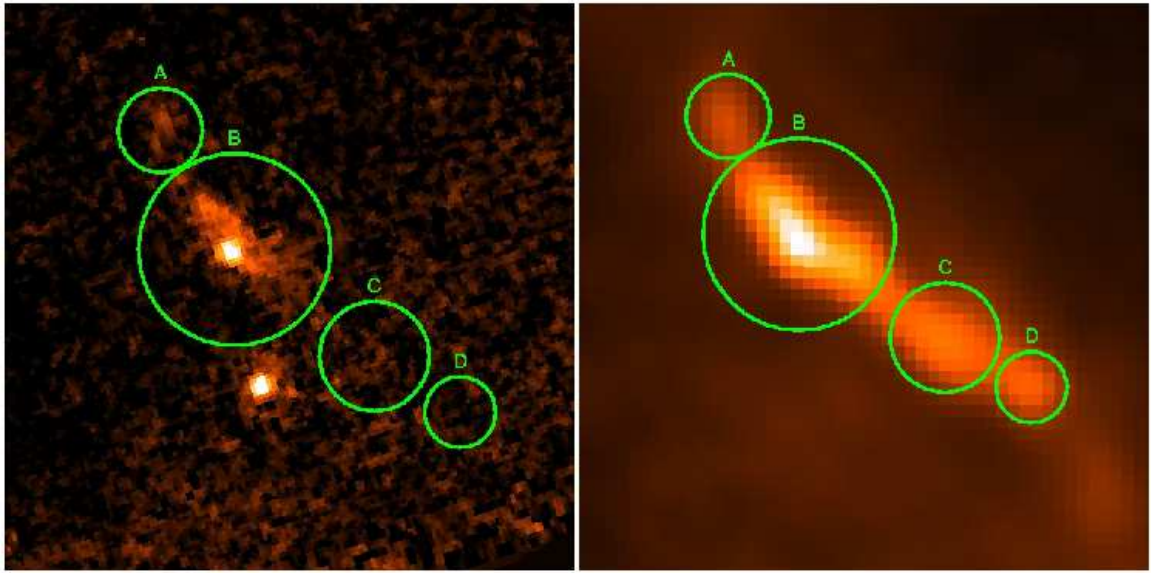}
\caption{The jet as  seen in the HST H-band image  (left panel) and in
the radio at 1.4 GHz (right). The projected angular size of the region
is $5\times 5 arcsec$.}
\label{radio_ir}
\end{figure*}

In Fig.~\ref{radio_ir} we show  a detail of  the jet as it  appears in
our IR HST image  (left panel) and  in the  VLA/MERLIN radio  image at
1.5GHz (right  panel).  In the  IR, the host  galaxy stellar component
has been subtracted using a simple  model obtained by fitting ellipses
to the galaxy surface brightness measured at different radii.

At least four emission regions, or ``knots'', can be identified in the
radio  jet: component ``B''  is the  brightest and  appears to  be the
largest at both frequencies.  Knots  ``A'' and ``B'' are also detected
in the IR image.

\begin{figure*}
\epsscale{1}
\plotone{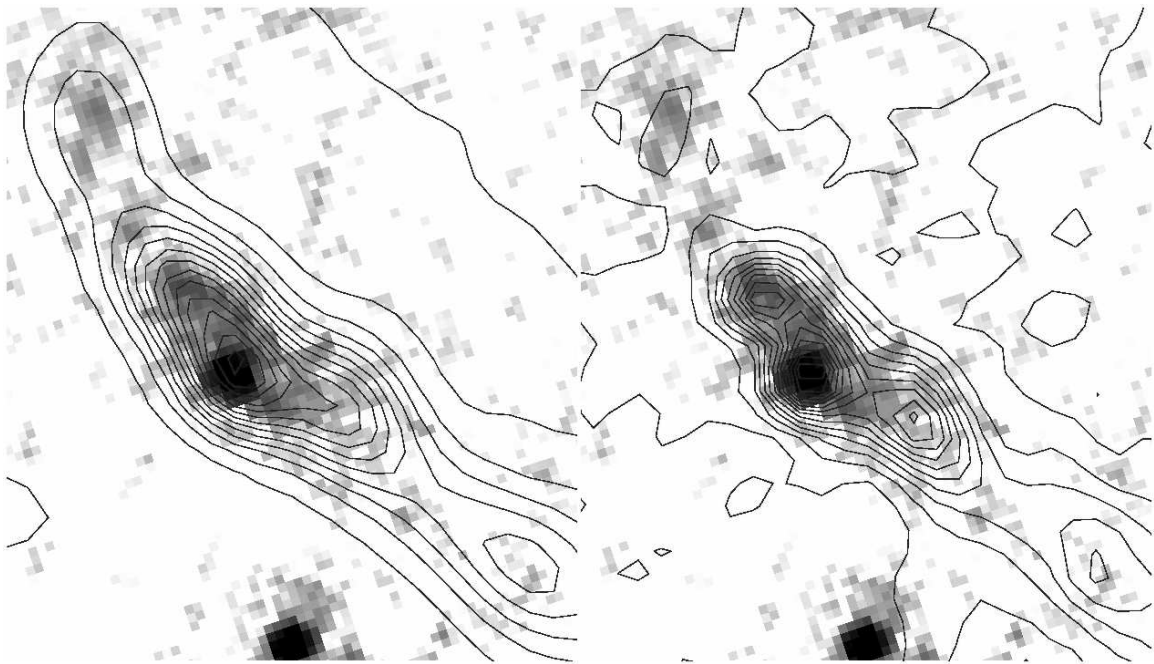}
\caption{Detail of the jet (component B is at the center of the image).
  The  radio  contours  at  1.5GHz  (left)  and  8.4GHz  (right) are
  overlaid  onto the  IR image.  The  projected angular  size of  the
  region is $2.5 \times 2.8 arcsec$.}
\label{jet_detail_cont}
\end{figure*}

Note  that component  ``B''  shows  a bright  point-like  knot in  the
infrared that has no radio counterpart at 1.5GHz. However the peaks of
emission in the infrared and in  the radio match quite well at 8.4 GHz
(the  data   are  described  in  Section   \ref{radiodata},  see  Fig.
\ref{jet_detail_cont}). Furthermore, in the radio maps at both 8.4 and
1.5GHz, the  contours are significantly  bent on the southern  side of
the jet,  where the peak of the  IR emission is located.   Thus we are
confident that the IR emission peak  is part of the jet itself, and it
is not produced by any background (or foreground) source.

We  measure the  flux of  component B  from our  HST H-band  image, by
extracting  the   counts  in  a  box   of  $1.7^{\prime\prime}  \times
1.2^{\prime\prime}$ and we measure  the background in different regions
around the jet.  We have performed aperture correction to estimate the
flux that is lost because of the presence of strong PSF wings. We have
performed  simulations using  a  synthetic PSF  obtained with  Tinytim
resulting in  a correction factor of  1.12 for the  extraction area we
use to measure  the flux of the jet.  The  flux, converted from counts
per second into physical units using the PHOTFLAM keyword in the image
header,  is  $6.0  \times 10^{-19}$  erg  cm$^{-2}$  s$^{-1}$
\AA$^{-1}$ with  estimated total error of  $\sim 8\%$. We  assume $A_V =
0.197$,  which translates  into  $A_H=0.034$ to  correct for  Galactic
absorption (as taken from NED).

Results of aperture  photometry for the   jet are summarized  in Table
\ref{tabphot}.

\section{Data at other wavelenghts: radio, K-band, optical, X-rays}
\label{otherobs}

\begin{deluxetable}{r  r}
\tablecolumns{2} 
\tablewidth{0pt}   \tablecaption{Jet fluxes for components A,B and C}
\tablehead{
\colhead{$\log \nu$} &   \colhead{$\log F_\nu$}\\
           \colhead{[Hz]}         &    \colhead{[erg cm$^-2$ s$^{-1}$ Hz$^{-1}$]}\\
}
\startdata
\cutinhead{Component A}
9.1761   &  -24.4560  \\
14.2730  &  -29.1348  \\
\cutinhead{Component B}
 9.1761    & -23.551  \\
 9.9243    & -24.032  \\
14.1347    & -27.979  \\
14.2730    & -28.223  \\ 
14.7367    & -28.858  \\ 
17.3838    & $<$-32.456  \\ 
\cutinhead{Component C}
9.1761   &  -24.0499   \\
14.2730  &  $<$-29.7620   \\
\enddata
\label{tabphot}
\end{deluxetable}

In order to derive  the SED of  the  jet we have  analyzed unpublished
data at different wavelengths,  from the radio  to the X-ray band. The
jet is detected in all of the observations we present here, except for
the X-ray band, for which we set an upper  limit.  In the following, we
briefly   describe    the data at   8.4    GHz  (VLA), K-band (GALILEO
telescope), optical (HST)  and X-rays  ({\it Chandra}), together  with
some details   on the reduction   procedures.  For  the data   at  all
wavelengths we use  the same extraction region  as  for the HST/NICMOS
image.

\subsection{Radio observations}
\label{radiodata}

As we  have already mentioned above,  we have obtained from  the NED a
VLA/MERLIN radio map  at 1.5 GHz.  We also  examined the NRAO archives
for  high-resolution ($<0.5$arcsec)  VLA observations  of  3C~401.  In
practice,  this  means observations  in  A  configuration at  $\lambda
\ltsim   6$  cm.   We  find   one   suitable  image   in  the   X-band
\citep[8.4GHz,][]{hardcastle97}.  The total on-source integration time
is  3580  seconds.   We  use   the  standard  {\sc  aips}  recipe  for
calibrating continuum  emission data,  using 3C~286 and  1926$+$611 as
our   flux  and  point-source   calibrators  respectively.    The  jet
morphology at  8.4 GHz  appears very similar  to that observed  in the
VLA/MERLIN map at 1.5 GHz.

\subsection{TNG K-band observations}

A K-band  image was obtained  at the 3.6m Italian  Telescopio Nazionale
Galileo, located at  La Palma, Canary Islands, Spain,  with the ARNICA
instrument \citep{lisi93}.  The observation  was part of our survey of
the complete sample of 3CR  sources with $z<0.3$ (P.I. Capetti), which
is described by  Marchesini et al. (in press).   The total integration
time is  1152s, split into 24 exposures  of 48s each to  allow for sky
subtraction. We reduce the data using the standard technique described
in Hunt  et al.  1994\footnote{Arcetri  technical report N.   4, 1994,
\url{http://www.arcetri.astro.it/science/irlab}}

The jet  is detected (at $4\sigma$) in  the  K-band Galileo image.  We
have extracted the flux  from the same region  as in the other images,
although  both the lower resolution  and the low  detected flux do not
allow us to disentangle the different jet components.

\subsection{Optical HST images}

3C~401 was  also observed with  HST/WFPC2 in two optical  bands, F702W
and F555W filters) as part of  programs 5476 and 6967 (P.I. Sparks for
both programs).  The presence of cosmic  rays in the region of the jet
in the  F702W image, for which  only one image was  taken, prevents us
from  estimating any  reliable upper  limit in  the R-band.   While no
detection can be  claimed in the F702W image, the  bright blob on knot
``B''  is clearly visible  in the  data taken  with the  F555W filter,
after smoothing the image with a 1-pixel $\sigma$ gaussian.

We thus  measure the flux of the  jet in the optical WFPC2/F555W image
over the same region of the sky as in  the IR NICMOS image.  We obtain
a  significant  (5$\sigma$)  excess  with  respect to   the background
level. The total error on the flux is $\sim 30\%$.

\subsection{Chandra data}


The  X-ray data   were  obtained  during two   different  observations
(Obs. Ids. 3083 and  4370, P.I. Reynolds) on  September 20th and  21st
2002, and are available on the Chandra public archive.

We  reduce the Chandra  data using the  Chandra data analysis software
CIAO  v3.0.2, with  the CALDB  version 2.25.  To   correct  for the QE
degradation issue we create a new ARF (effective area files) using the
file {\tt acisD1999-08-13contamN0003.fits} released  with the CALDB  version
2.26. We reprocess all the data from level  1 to level 2 with standard
procedures.  No background flares   are present during the   observing
time.  Thus, merging the two  datasets, we obtain  an event file with
total exposure time  of 47.52 Ks.  We extracted   the spectrum in  the
region of the IR jet and we measure the background in a nearby region.

We  measure  a  $3\sigma$  upper  limit  to  the  jet  count  rate  of
$3.59\times 10^{-5}$ cts/s.   Since the flux is too  low to derive any
reasonable fit to the spectrum, we use the Redistribution Matrix (RMF)
and Auxiliary  Response File (ARF)  created by CIAO to  reproduce with
Xspec  the  observed upper  limit  net  count  rate with  an  absorbed
power-law  model.   The  value  of  $N_H$ is  fixed  to  the  galactic
absorption ($6.77\times 10^{20}$ cm$^{-2}$)  and the power-law index to
the value  of $\alpha=1$.  Note that a  different choice  for $\alpha$
would  not affect  significantly the  normalization of  the power-law.
The  normalization of the  power-law results  in $5.98  \times 10^{-8}$
photons  cm$^{2}$  s$^{-1}$  keV$^{-1}$  at 1  keV,  corresponding  to
$F_{1keV} < 0.35$ nJy (3$\sigma$ upper limit).

\section{Spectral energy distribution and emission model}
\label{sedsect}

\begin{figure}
\epsscale{1}
\plotone{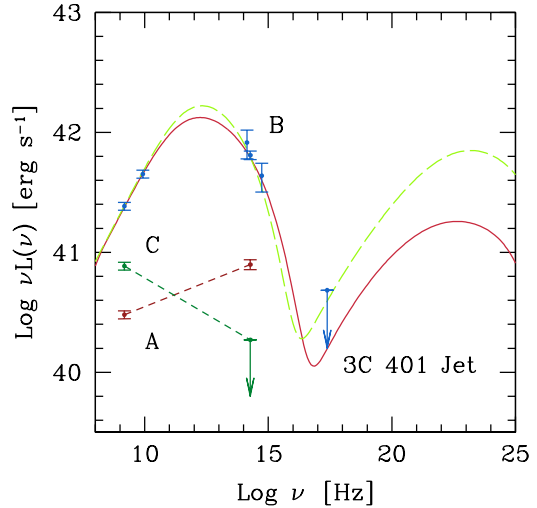}
\caption{SED and two  models for component ``B''. The  model parameters are
given in Table~\ref{modpar}. The dashed line corresponds to model 1 in
the table, the solid line corresponds to model 2. We also report radio
(1.5GHz)  and infrared (1.6$\mu$m)  measurements for  components ``A''
and ``C''.``}
\label{sed}
\end{figure}

In  Fig.   \ref{sed}  we  show  the spectral  energy  distribution  of
component ``B'' of the jet.  Note  that although the jet is very faint
and only  marginally detected in  the optical, the  optical data-point
sets a crucial  constraint on the shape of the jet  SED.  The slope is
flat ($\alpha = 0.64 \pm 0.06$)  in the radio, and  changes to
$\alpha > 1$ in the spectral region between the radio and the IR band.

We have  also measured the fluxes  for components ``A''  and ``C'', in
order  to investigate  any spectral  variation along  the  jet.  While
``A'' is  detected both in the  radio and at 1.6$\mu$m,  ``C'' is only
visible  at radio  wavelengths.   For the  latter  component, we  have
measured a 3$\sigma$ upper limit  from our 1.6$\mu$m HST image.  Both A
and B have a broad-band  spectral index $\alpha_{1.5GHz-IR} = 0.92 \pm
0.01$ while  C is significantly steeper  ($\alpha_{1.5GHz-IR} > 1.1$).
For  component   B,  we   can  also  set   useful  lower   limits  for
$\alpha_{IR-X}\gta 1.4 $ and $\alpha_{1.5GHz-X}\gta 1.09 $, that we use
in the  following section  to compare the  properties of our  jet with
those of other known X-ray and radio jets.

The continuous and the dashed lines in Fig.\ref{sed} correspond to two
possible  one-zone  synchrotron  self-Compton (SSC)  emission  models,
obtained with different sets  of physical parameters, as summarized in
Table  \ref{modpar}.   The volume  of  the  emitting  region has  been
constrained  from the  observed  size of  the  IR jet  (assumed to  be
cylindrical) and the  spectral slope in the radio  band constrains the
electron injected  spectrum. For model  1 (dashed line), the  value of
the magnetic  field is close  to equipartition (within a  factor $\sim
3$), while  for model 2  (solid line) we  fixed the beaming  factor to
unity and the  magnetic field is derived from fitting  the SED, as for
the remaining  parameters.  In the model computation,  the geometry of
the source  is assumed to  be spherical, and electrons  are constantly
injected to  reach equilibrium.  Synchrotron and SSC  losses are taken
into  account, while  electrons  can  escape from  the  source with  a
typical  time-scale given  by  $t_{esc}$ \citep[see][]{chiabgg}.   The
minimum energy  of the  injected electron distribution,  together with
the escape time, are set in  order not to overpredict the emission
of   the  lower   energy   electrons  in   both   the  synchrotron   and
inverse-Compton components (i.e. the flat regions of the SED, in the 
radio and X-ray).

\begin{deluxetable}{l  l}
\tablewidth{0pt}   \tablecaption{Model   parameters for component B}
\tablehead{\colhead{Model 1} &   \colhead{Model 2}}
\startdata
$R  = 5.0 \times 10^{21}$ cm               & $R  = 5.0 \times 10^{21}$ cm                  \\
$L_{inj}= 1.1 \times 10^{45}$ erg s$^{-1}$ & $L_{inj}= 3.7 \times 10^{44}$ erg s$^{-1}$    \\
$\gamma_{min}= 500 $                       & $\gamma_{min}= 10 $                           \\ 
$\gamma_{max}= 3.5 \times 10^{6}$	   & $\gamma_{max}= 3.5 \times 10^{6}$		   \\
$p= 2.2 $                                  & $p= 2.2 $                                     \\ 
$B =  1.0 \times 10^{-4}$ G                & $B =  1.5 \times 10^{-5}$ G                   \\
$\delta =0.5$                    	   & $\delta =1$                    		   \\
$t_{esc} = 6 R/c$             		   & $t_{esc} = 80 R/c$             		   \\
\hline
\enddata
\label{modpar}
\end{deluxetable}

\section{Discussion and conclusions}
\label{conclu}

An important point in understanding  the nature of the jet emission is
to address the following question: what is the reason for the observed
brightening of  the jet  at a projected  distance from the  nucleus of
$\sim 20$kpc?  The jet in 3C~401  appears to be brighter in the region
in  which  bending occurs.   This  can  be  explained by  relativistic
beaming, or by  the presence of shocks which  accelerate the electrons
along the jet, or by a combination  of the two. While in FR~I jets the
jet  flow most  likely slows  down to  sub-sonic  and sub-relativistic
velocities on scales smaller than $\sim 1$kpc, there are clues that in
FR~II the  jets are still relativistic  all the way  to the hot-spots.
Radio observations give us a  clue that relativistic bulk motion plays
a  role in  the jet  of  3C~401.  In  fact, \citet{hardcastle98}  have
measured that  in a region close to  the radio core, but  still on the
scale of  kpc, the jet-to-counterjet ratio  is $R_{j-cj}>5$.  Although
this limit does not allow us  to set any firm constraint to either the
viewing angle to the jet or  to the bulk Lorentz factor of the ejected
material, it is an indication that the jet is relativistic at least on
those scales.  Furthermore, it is widely believed that the presence of
hot-spots in the  lobes of the radio emission is  a signature that the
jets in FR~II are still supersonic in the region where they impact the
ambient  medium,   and  possibly  also   relativistic  \citep{markos}.
Therefore, we believe that the  most likely interpretation is that the
enhancement  of the  flux is  partly or  entirely due  to relativistic
beaming.   However, with  the present  observations we  cannot totally
rule  out that  part  of the  jet  enhancement occurs  because of  the
presence of  shocks in  the jet.  If  this is  the case, we  expect to
observe a progressive steepening  of the IR-to-UV spectral slope along
the jet as a result of electron aging.

We can estimate the amount  of beaming involved, by comparing the flux
of the jet  in the detected and undetected region.   From the IR image
we  derive that  the jet  is enhanced  by a  factor of  $\gta  20$, as
estimated from the ratio between the  flux of blob ``B'' and the upper
limit  derived on  a region  where  the jet  is undetected.   Assuming
$\delta_B   =1$   as  obtained   from   model   2,   we  obtain   that
$(\delta_{B}/\delta_{und})^4  \gta 20$  ($\delta_{und}$ refers  to the
region  where the jet  is undetected),  thus $\delta_{und}  \lta 0.5$.
For $\Gamma=5$  this implies  a change in  the viewing angle  of $\gta
35^{\circ}$   (from   $\theta  \gta   60^{\circ}$   to  $\theta   \sim
35^{\circ}$).   Note that  under  the above  assumptions, the  Lorentz
factor   cannot  be   lower  than   2,   since  for   $\Gamma  <   2$,
$\delta(90^\circ)  > 0.5$.  Higher  values of  $\Gamma$ imply  an even
smaller change  in the  direction of the  jet (e.g. from  $\theta \gta
30^\circ$ to $\theta  \sim 25^\circ$, for $\Gamma =  10$) but there is
no  strong evidence  for  such  a fast  relativistic  motion from  the
observations. A similar  argument holds for model 1,  in which we have
$\delta = 0.5$.

The  presence of  relativistic motion  on large  scales FR~II  jets is
indeed  not surprising.   However,  the evidence  for  that is  mainly
indirect. One  of the most  convincing arguments is derived  from the
high Lorentz factors needed to enhance the cosmic microwave background
radiation field which is up-scattered by the relativistic particles in
the jet to produce the observed X--ray flux \citep{tavecchio00,cgc01}.
But contrary  to most (if not all)  other ``optically-bright'' jets,
in  the case  of 3C~401  we do  not detect  any X-rays  emission.  The
reason for the lack of significant high-energy flux might be i)
``intrinsic'' or ii) ``extrinsic''.

i) The  particle density in the  comoving frame is in  fact quite low.
From  the  model  fitting,  we   derive  a  value  for  the  so-called
``compactness parameter'' $\ell_{mod 1} = L_{inj_1} \sigma_T/R m_e c^3
= 6.0 \times 10^{-6}$ and $\ell_{mod 2} = L_{inj_2} \sigma_T/R m_e c^3
= 2.0 \times  10^{-6}$ (e.g. Guilbert, Fabian, Rees  1983), which is a
measure of  the ``density'' of  the source ($L_{inj}$ is  the injected
luminosity in particles, $\sigma_T$  in the Thomson cross section, $R$
is  the size  of the  source, $m_e$  is the  electron mass).   Thus the
synchrotron emission  peak dominates with  respect to the  high energy
(SSC)  peak.  In  this  framework, the  source  is thus 
dominated by synchrotron emission, while the SSC 
radiation is intrinsically low.

ii) When compared to the  magnetic energy density, the external photon
field  is  low  as  well.   In  fact, we  find  that,  for  $z=0.2$,
$U^{\prime}_{CMB}/U^{\prime}_B  \sim 3 \times  10^{-3}$ (model  1) and
$0.1$ (model 2), where $U^{\prime}_{CMB}$ is the energy density of the
cosmic microwave  background calculated in  the rest frame of  the jet,
assuming   $\Gamma  =  5$   and  a   viewing  angle   of  $30^{\circ}$
\citep{harriskrav}. Thus, synchrotron  radiation is again the dominant
radiation mechanism.

As  pointed out by  \citet{cgc01}, a  third source  of photons  for IC
scattering may reside  in the ``beam'' from the  central quasar hosted
by the nucleus of the galaxy. If the jet axis forms a large angle with
the line-of-sight, the central quasar photons, that are preferentially
emitted  in the  direction of  the jet  axis, can  be  up-scattered by
relativistic electrons in the jet. Then, if the jet has a slower shear
layer on the  kpc scales, \citep[as it has been  suggested in order to
explain the limb-brightnened  appearence of the jet in  e.g. the FR~II
radio galaxy 3C~353][]{swain}, the  IC radiation is not heavily beamed
away from the observer, and  it can be detected.  However, 3C~401 does
not appear to  harbor any powerful quasar.  This  can be inferred both
from its  optical spectral characteristics  and from the  nuclear SED:
3C~401 is a  narrow-lined Low Excitation Galaxy (LEG),  which does not
appear to require  a high ionizing nuclear flux, as  it is for objects
with  high  excitation  emission  lines  (QSO  and  narrow-lined  High
Excitation galaxies, HEG).  Furthermore,  the object does not show any
strong central thermal emission  of associated with the active nucleus
both in the optical \citep{pap4} and  in the IR.  This is clearly seen
in our HST  images in which only a faint  central nuclear component is
observed  (the  nuclear properties  of  the  complete  sample will  be
discussed in a forthcoming paper, Chiaberge et al., in preparation).

A further characteristic  of the jet SED  is worth discussing. In order
not  to produce significant X-ray  flux, the electron distribution
must have a cut-off  at energies confined between  the optical and the
X-ray band.  As derived from the SED modeling, for a magnetic field of
$B=10^{-4}$ G,   the cut-off energy should  be   located at  a Lorentz
factor of $\gamma   \sim$  a few   $10^6$.  Typical maximum   electron
energies  in synchrotron-dominated FR~I  jets are significantly higher
(up to  $\gamma \sim$ a  few $10^7-10^{8}$),  while \citet{sambruna04}
have shown that jets in powerful FR~II/quasars have $\gamma_{max} \sim
5\times 10^5- 10^6$.

Thus,  on the  one  hand, our  3C~401  jet shows  properties that  are
intermediate  between the  two classes  of synchrotron  dominated, low
power  FR~I jets  and  IC-dominated, high  power  FR~II/QSO: its  bulk
velocity on the kpc-scale and its electron distribution are typical of
high power jets,  while the dominance of synchrotron  radiation in the
overall SED is typical of low power jets. On the other hand, the broad
band  spectral indices  $\alpha_{ro} \sim  0.9$ and  $\alpha_{ox} \gta
1.4$ of knot  ``B''' are atypical (see Fig.   4 in \citet{sambruna04})
as it  would lie in  an empty region of  the $\alpha_{ro}-\alpha_{ox}$
plane.

An  intriguing   consequence  is  that  there  may   be  a  continuous
``sequence'' in extended X-ray  jets, somehow similar to that observed
in   jets  on  a   much  smaller   scale  in   the  case   of  blazars
\citep{fossati98,gg98}.   In this framework,  our jet  would represent
the prototype of  the new ``intermediate class'', and  their study may
shed light on the physical  differences between the various jets. From
the point of view of  the observations, the most important property of
such  intermediate  jets is  that  their  SED  peaks in  the  infrared
region.

However, we  must stress that the  jet should be studied  in much more
detail before  drawing conclusions.  A  crucial point is to  produce a
spatially  resolved spectral  energy  distribution of  knot ``B'',  in
order to disentangle the bright  ``spot'' seen in the IR (and possibly
at 8.4 GHz) from the rest of  the jet. If that ``spot'' is a region of
the jet where  electrons are being accelerated, for  example where the
jet impacts the ISM/IGM, then its SED might be substantially different
from the rest of the jet,  thus affecting our conclusions based on the
``spatially integrated'' SED shown  in this paper.  In particular, the
region to  be explored is from the  IR K-band to the  near UV, through
deep  high-resolution  observations that  only  the  HST can  provide.
Another  important  issue  to  be  investigated  through  deeper  high
resolution  images is  the change  in the  spectral slope  between the
different components of  the jet. This may shed  light on the possible
role of shocks in observed jet brightening and change of direction.

\acknowledgments We are grateful to Eddie Bergeron and Juan Madrid for
helping us to  reduce the NICMOS data. We  thank Markos Georganopoulos
for  insightful  discussions.    MC  acknowledges  the  STScI  Visitor
Program.   We thank J.~P. Leahy for  providing the  1.5 GHz  radio map,
which is available at http://www.jb.man.ac.uk/atlas.

This  paper is  based on
observations obtained at the  Space Telescope Science Institute, which
is  operated  by  the  Association  of Universities  for  Research  in
Astronomy,  Incorporated,  under  NASA  contract  NAS  5-26555.   This
research has  made use of  the NASA/IPAC Extragalactic  Database (NED)
which  is  operated  by  the  Jet  Propulsion  Laboratory,  California
Institute of Technology, under  contract with the National Aeronautics
and Space Administration.

\end{document}